\documentclass[amsmath,aps,showpacs,a4paper,10pt]{revtex4}

 \usepackage{epsf}
 \usepackage{graphicx}    

\begin{document}

 \newcommand{\be}[1]{\begin{equation}\label{#1}}
 \newcommand{\ee}{\end{equation}}
 \newcommand{\bea}{\begin{eqnarray}}
 \newcommand{\eea}{\end{eqnarray}}
 \def\disp{\displaystyle}

 \def\gsim{ \lower .75ex \hbox{$\sim$} \llap{\raise .27ex \hbox{$>$}} }
 \def\lsim{ \lower .75ex \hbox{$\sim$} \llap{\raise .27ex \hbox{$<$}} }

 \begin{titlepage}

 \begin{flushright}
 arXiv:1012.0883
 \end{flushright}

 \title{\Large \bf Dark Energy Cosmology with
 the~Alternative~Cosmic~Microwave~Background~Data}

 \author{Hao~Wei\,}
 \email[\,email address:\ ]{haowei@bit.edu.cn}
 \affiliation{Department of Physics, Beijing Institute
 of Technology, Beijing 100081, China}

 \begin{abstract}\vspace{1cm}
 \centerline{\bf ABSTRACT}\vspace{2mm}
 Recently, in a series of works by Liu and Li (L\&L),
 they claimed that there exists a timing asynchrony of
 $-25.6\,$ms between the spacecraft attitude and radiometer
 output timestamps in the original raw WMAP time-ordered data
 (TOD). L\&L reprocessed the WMAP data while the aforementioned
 timing asynchrony has been corrected, and they obtained an
 alternative CMB map in which the quadrupole dropped to nearly
 zero. In the present work, we try to see the implications to
 dark energy cosmology if L\&L are right. While L\&L claimed
 that there is a bug in the WMAP pipeline which leads to
 significantly different cosmological parameters,
 an interesting question naturally arises, namely, how robust
 is the current dark energy cosmology with respect
 to systematic errors and bugs? So, in this work, we adopt the
 alternative CMB data of L\&L as a strawman to study the
 robustness of dark energy predictions.
 \end{abstract}

 \pacs{98.80.Es, 95.36.+x, 98.80.Cq, 98.70.Vc}

 \maketitle

 \end{titlepage}

 \renewcommand{\baselinestretch}{1.0}


\section{Introduction}\label{sec1}

Dark energy cosmology has been one of the most active fields
 in astronomy and physics, since the exciting discovery of
 current accelerated expansion of our universe~\cite{r1}.
 The first evidence came from the observation of Type~Ia
 supernovae (SNIa) in 1998~\cite{r2}. Five years later,
 cosmology entered the so-called ``precision era'' in 2003
 when the first year Wilkinson Microwave Anisotropy Probe
 (WMAP) observations of the cosmic microwave background (CMB)
 had been released~\cite{r3}. Up to now, the CMB observation
 is still a very powerful probe for cosmology, and the CMB
 data from WMAP mission provide the most important basis.

However, in the passed years, some unusual phenomena have been
 found from the CMB data released by WMAP team. Remarkably,
 among these unusual phenomena, it is claimed that there exists
 a preferred direction in the CMB temperature map (known as the
 ``Axis of Evil'' in the literature)~\cite{r4}. In fact, there
 are two different approaches to deal with this problem. The
 first one is to admit that this phenomenon is an observational
 fact and try to explain it in the cosmological theories (see
 e.g.~\cite{r5,r6,r7} and references therein). The second
 approach is to consider instead that this phenomenon might be
 an artifact due to the observational systematics. For example,
 in a series of works by Liu and Li (L\&L)~\cite{r8,r9,r10},
 they claimed that there exists a timing asynchrony of
 $-25.6\,$ms between the spacecraft attitude and radiometer
 output timestamps in the original raw WMAP time-ordered data
 (TOD). If one fixed this problem, the most of CMB quadrupole
 component disappear. It is worth noting that recently their
 findings has been confirmed by several independent authors
 (see e.g.~\cite{r11,r12,r13}). In~\cite{r8}, L\&L reprocessed
 the WMAP data while the aforementioned timing asynchrony has
 been corrected, and they obtained an alternative CMB map in
 which the quadrupole dropped to nearly zero. In addition,
 using the alternative CMB data, they constrained the cosmological
 parameters of the standard flat $\Lambda$CDM model, and
 found that these cosmological parameters have been changed
 notably. For convenience, we reproduce their main results in
 Table~\ref{tab1}. It is easy to see that the fractional
 matter density $\Omega_{m0}=\Omega_{b0}+\Omega_{c0}$ of
 L\&L~\cite{r8} is considerably larger than the one of
 WMAP~\cite{r14,r15}.\\[-1mm]


 \begin{table}[htbp]
 \begin{center}
 \begin{tabular}{cccc} \hline\hline\\[-3mm]
 Description & Symbol & WMAP~\cite{r14,r15} & L\&L~\cite{r8}
 \\[1.2mm] \hline \\[-3mm]
 ~Hubble constant (km/s/Mpc)~~~~~ & $H_0$ & $71.9^{+2.6}_{-2.7}$
 & $71.0\pm 2.7$ \\
 Baryon density & $\Omega_{b0}$ & ~~~~~$0.0441\pm 0.0030$~~~~~~
 & $0.052\pm 0.0030$~ \\
 Cold dark matter density & $\Omega_{c0}$ & $0.214\pm 0.027$
 & $0.270\pm 0.027$ \\
 Dark energy density & $\Omega_{\Lambda 0}$ & $0.742\pm 0.030$
 & $0.678\pm 0.030$ \\
 Fluc. Ampl. at $8h^{-1}\,$Mpc & $\sigma_8$ & $0.796\pm 0.036$
 & $0.921\pm 0.036$ \\
 Scalar spectral index & $n_s$ & $0.963^{+0.014}_{-0.015}$
 & $0.957\pm 0.015$ \\
 Reionization optical depth & $\tau$ & $0.087\pm 0.017$
 & $0.109\pm 0.017$ \\[1.2mm]
 \hline\hline
 \end{tabular}
 \end{center}
 \caption{\label{tab1} The cosmological constraints on the
 standard flat $\Lambda$CDM model, reproduced from~\cite{r8}.}
 \end{table}


To our knowledge, there is still controversy about the findings
 of L\&L in the community. In the present work, we try to be
 neutral as much as possible. Instead of discussing the detailed
 data process of WMAP mission (as in the works by
 L\&L~\cite{r8,r9,r10} or other authors~\cite{r11,r12,r13}), in
 this work we would like to see the implications to dark energy
 cosmology if L\&L are right. While L\&L claimed that there is
 a bug in the WMAP pipeline which leads to significantly
 different cosmological parameters, an interesting question
 naturally arises, namely, how robust is the current dark
 energy cosmology with respect to systematic errors and bugs?
 So, in this work, we adopt the alternative CMB data of L\&L as
 a strawman to study the robustness of dark energy predictions.

As is well known, using the full CMB data to perform a global
 fitting consumes a large amount of computation time and power.
 As a good alternative, one can instead use the shift parameter
 $R$ from the CMB data, which has been considered extensively
 in the literature. It is argued that the shift parameter $R$
 is model-independent, and it contains the main information of
 the CMB data~\cite{r16}. So, in Sec.~\ref{sec2} we firstly
 derive the corresponding shift parameter $R$ from the
 alternative CMB data of L\&L~\cite{r8}. In addition, we
 briefly introduce the other observational data, such as SNIa
 and the baryon acoustic oscillation (BAO), which are also used
 in the present work. In Sec.~\ref{sec3}, we perform the data
 analysis by using the cosmological observations introduced in
 Sec.~\ref{sec2}. In particular, we discuss the tension between
 CMB and SNIa in Sec.~\ref{subsec3}; we study the age problem
 in dark energy models in Sec~\ref{subsec4}; and
 the cosmological constraints on dark energy models are
 considered in Sec.~\ref{subsec5}. Finally, the conclusion and
 discussions are given in Sec.~\ref{sec6}.


\section{Observational data}\label{sec2}


\subsection{Shift parameter $R$ from the alternative CMB data}\label{sec2a}

As mentioned above, using the shift parameter $R$ from the
 CMB data is a good approach to fit the cosmological models.
 As is well known, the shift parameter $R$ of the CMB data
 is defined by~\cite{r17} (see also~\cite{r14,r15,r16})
 \be{eq1}
 R\equiv\Omega_{m0}^{1/2}\int_0^{z_\ast}
 \frac{d\tilde{z}}{E(\tilde{z})}\,,
 \ee
 where $\Omega_{m0}$ is the present fractional energy density
 of pressureless matter; $E\equiv H/H_0$ in which
 $H\equiv\dot{a}/a$ is the Hubble parameter; $a=(1+z)^{-1}$ is
 the scale factor (we have set $a_0=1$; the subscript ``0''
 indicates the present value of corresponding quantity; $z$ is
 the redshift); a dot denotes the derivative with respect to
 cosmic time $t$; the redshift of recombination $z_\ast$ is
 given by~\cite{r18} (see also~\cite{r14,r15})
 \be{eq2}
 z_\ast=1048\left[1+0.00124\left(\Omega_{b0}h^2\right)^{-0.738}
 \right]\left[1+g_1\left(\Omega_{m0}h^2\right)^{g_2}\right],
 \ee
 where $\Omega_{m0}=\Omega_{b0}+\Omega_{c0}$, and
 $\Omega_{b0}$, $\Omega_{c0}$ are the present fractional energy
 densities of baryon and cold dark matter, respectively; $h$ is
 the Hubble constant $H_0$ in units of $100\,$km/s/Mpc; and
 \be{eq3}
 g_1=\frac{0.0783\left(\Omega_{b0}h^2\right)^{-0.238}}
 {1+39.5\left(\Omega_{b0}h^2\right)^{0.763}}\,,~~~~~~~~~~
 g_2=\frac{0.560}{1+21.1\left(\Omega_{b0}h^2\right)^{1.81}}\,.
 \ee
 The shift parameter $R$ relates the angular diameter distance
 to the last scattering surface, the comoving size of the
 sound horizon at $z_\ast$ and the angular scale of the first
 acoustic peak in CMB power spectrum of temperature
 fluctuations~\cite{r16,r17}.

In principle, one should obtain $z_\ast$ and $R$ from the full
 (alternative) CMB data. However, this is a hard work consuming a
 large amount of computation time and power. On the other hand,
 L\&L have not published their full alternative CMB data in which
 the timing asynchrony of $-25.6\,$ms between the spacecraft
 attitude and radiometer output timestamps in the original raw
 WMAP time-ordered data (TOD) has been corrected. Therefore,
 in this work we use instead the Monte Carlo method~\cite{r19}.
 We choose the standard flat $\Lambda$CDM model to be the
 fiducial model, since its cosmological parameters have been
 constrained by both WMAP~\cite{r14,r15} and L\&L~\cite{r8}
 (see Table~\ref{tab1}). We generate the Gaussian distributions
 for $\Omega_{b0}$, $\Omega_{c0}$ and $h$ from their best-fit
 parameters and the corresponding $1\sigma$ uncertainties given
 in Table~\ref{tab1}. Then, we randomly sample the parameters
 $\Omega_{b0}$, $\Omega_{c0}$ and $h$ from their corresponding
 Gaussian distribution for $N_{\rm mc}$ times. For each
 $\{\Omega_{b0},\,\Omega_{c0},\,h\}$, we can obtain the
 corresponding $z_\ast$ and $R$ from Eqs.~(\ref{eq2}) and
 (\ref{eq1}), respectively. Finally, we can determine the means
 and the corresponding $1\sigma$ uncertainties for $z_\ast$ and
 $R$ from these $N_{\rm mc}$ samples, respectively. Notice that
 in this work we have done $N_{\rm mc}=10^6$ samplings. At
 first, we check this method by calculating $z_\ast$ and $R$
 from the WMAP data, and find that our results are well
 consistent with the ones given by WMAP team~\cite{r14,r15}.
 Then, we turn to the alternative CMB data of L\&L~\cite{r8}, and
 finally find that the corresponding redshift of recombination
 $z_\ast$ is given by
 \be{eq4}
 z_\ast=1088.610\pm 2.373\,,
 \ee
 and the shift parameter $R$ is given by
 \be{eq5}
 R=1.761\pm 0.014\,.
 \ee
 Obviously, the shift parameter $R$ from L\&L's alternative CMB
 data is fairly larger than the one of WMAP~\cite{r14,r15}.
 It is argued that the shift parameter $R$ is
 model-independent, and it contains the main information of
 the CMB data~\cite{r16}. Therefore, in the literature the
 shift parameter $R$ has been used extensively to constrain
 cosmological models. For the alternative CMB data of
 L\&L~\cite{r8}, the corresponding $\chi^2$ from the shift
 parameter $R$ reads
 \be{eq6}
 \chi^2_{\rm CMB}=\left(\frac{R-1.761}{0.014}\right)^2.
 \ee


\subsection{SNIa and BAO data}\label{sec2b}

In addition to the CMB data, we also consider the observations
 of SNIa and BAO. The data points of the latest 557 Union2 SNIa
 compiled in~\cite{r20} are given in terms of the distance
 modulus $\mu_{obs}(z_i)$. On the other hand, the theoretical
 distance modulus is defined as
 \be{eq7}
 \mu_{th}(z_i)\equiv 5\log_{10}D_L(z_i)+\mu_0\,,
 \ee
 where $\mu_0\equiv 42.38-5\log_{10}h$ and
 \be{eq8}
 D_L(z)=(1+z)\int_0^z \frac{d\tilde{z}}{E(\tilde{z};{\bf p})}\,,
 \ee
 in which ${\bf p}$ denotes the model parameters. The
 corresponding $\chi^2$ from the 557 Union2 SNIa reads
 \be{eq9}
 \chi^2_{\mu}({\bf p})=\sum\limits_{i}\frac{\left[
 \mu_{obs}(z_i)-\mu_{th}(z_i)\right]^2}{\sigma^2(z_i)}\,,
 \ee
 where $\sigma$ is the corresponding $1\sigma$ error. The parameter
 $\mu_0$ is a nuisance parameter but it is independent of the data
 points. One can perform an uniform marginalization over $\mu_0$.
 However, there is an alternative way. Following~\cite{r21,r22,r28},
 the minimization with respect to $\mu_0$ can be made by expanding
 the $\chi^2_{\mu}$ of Eq.~(\ref{eq9}) with respect to $\mu_0$ as
 \be{eq10}
 \chi^2_{\mu}({\bf p})=\tilde{A}-2\mu_0\tilde{B}+\mu_0^2\tilde{C}\,,
 \ee
 where
 $$\tilde{A}({\bf p})=\sum\limits_{i}\frac{\left[\mu_{obs}(z_i)
 -\mu_{th}(z_i;\mu_0=0,{\bf p})\right]^2}
 {\sigma_{\mu_{obs}}^2(z_i)}\,,$$
 $$\tilde{B}({\bf p})=\sum\limits_{i}\frac{\mu_{obs}(z_i)
 -\mu_{th}(z_i;\mu_0=0,{\bf p})}{\sigma_{\mu_{obs}}^2(z_i)}\,,
 ~~~~~~~~~~
 \tilde{C}=\sum\limits_{i}\frac{1}{\sigma_{\mu_{obs}}^2(z_i)}\,.$$
 Eq.~(\ref{eq10}) has a minimum for
 $\mu_0=\tilde{B}/\tilde{C}$ at
 \be{eq11}
 \tilde{\chi}^2_{\mu}({\bf p})=
 \tilde{A}({\bf p})-\frac{\tilde{B}({\bf p})^2}{\tilde{C}}\,.
 \ee
 Since $\chi^2_{\mu,\,min}=\tilde{\chi}^2_{\mu,\,min}$
 obviously, we can instead minimize $\tilde{\chi}^2_{\mu}$
 which is independent of $\mu_0$. It is worth noting that the
 corresponding $h$ can be determined by $\mu_0=\tilde{B}/\tilde{C}$
 for the best-fit parameters.

Similar to the case of shift parameter $R$, one can use the
 the distance parameter $A$ of the measurement of the BAO peak
 in the distribution of SDSS luminous red galaxies~\cite{r23},
 which is also model-independent and contains the main
 information of the observation of BAO. The distance parameter
 $A$ is defined by~\cite{r23}
 \be{eq12}
 A\equiv\Omega_{m0}^{1/2}E(z_b)^{-1/3}\left[\frac{1}{z_b}
 \int_0^{z_b}\frac{d\tilde{z}}{E(\tilde{z})}\right]^{2/3},
 \ee
 where $z_b=0.35$. In~\cite{r24}, the value of $A$ has been
 determined to be $0.469\,(n_s/0.98)^{-0.35}\pm 0.017$. Here
 the scalar spectral index $n_s$ is taken to be $0.957$ from
 L\&L~\cite{r8} (see also Table~\ref{tab1}). The corresponding
 $\chi^2_{\rm BAO}=(A-A_{\rm obs})^2/\sigma_{\rm A}^2$.

The best-fit model parameters are determined by minimizing the
 corresponding $\chi^2$. As in~\cite{r25,r26,r76}, the $68.3\%$
 confidence level (C.L.) is determined by
 $\Delta\chi^2\equiv\chi^2-\chi^2_{min}\leq 1.0$, $2.3$ and
 $3.53$ for $n_p=1$, $2$ and~$3$, respectively, where $n_p$ is
 the number of free model parameters. Similarly, the $95.4\%$
 C.L. is determined by
 $\Delta\chi^2\equiv\chi^2-\chi^2_{min}\leq 4.0$, $6.17$ and
 $8.02$ for $n_p=1$, $2$ and $3$, respectively.\\


 \begin{table}[tbhp]
 \begin{center}
 \begin{tabular}{lcccc} \hline\hline \\[-3.5mm]
 ~~ Observation & $\chi^2_{min}$ & $\Omega_{m0}$ & $w_0$
 & $w_a$ \\[0.8mm] \hline \\[-3.5mm]
 ~~ SNIa & ~~~~ 541.430 ~~~~ & ~~~~ 0.420 ~~~~
 & ~~~~ $-0.863$ ~~~~ & ~~ $-5.490$ ~~ \\
 ~~ SNIa+$A$ & 542.636 & 0.278 & $-1.007$ & $-0.105$ \\
 ~~ SNIa+$R$ & 542.523 & 0.295 & $-1.021$ & $-0.314$ \\
 ~~ SNIa+$A$+$R$ ~~~ & 542.936 & 0.288 & $-0.969$ & $-0.529$
 \\[0.8mm] \hline\hline
 \end{tabular}
 \end{center}
 \caption{\label{tab2} The $\chi^2_{min}$ and the best-fit
 values of $\Omega_{m0}$, $w_0$ and $w_a$ for the various
 observations.}
 \end{table}


\vspace{-5mm}  


 \begin{center}
 \begin{figure}[tbhp]
 \centering
 \includegraphics[width=1.0\textwidth]{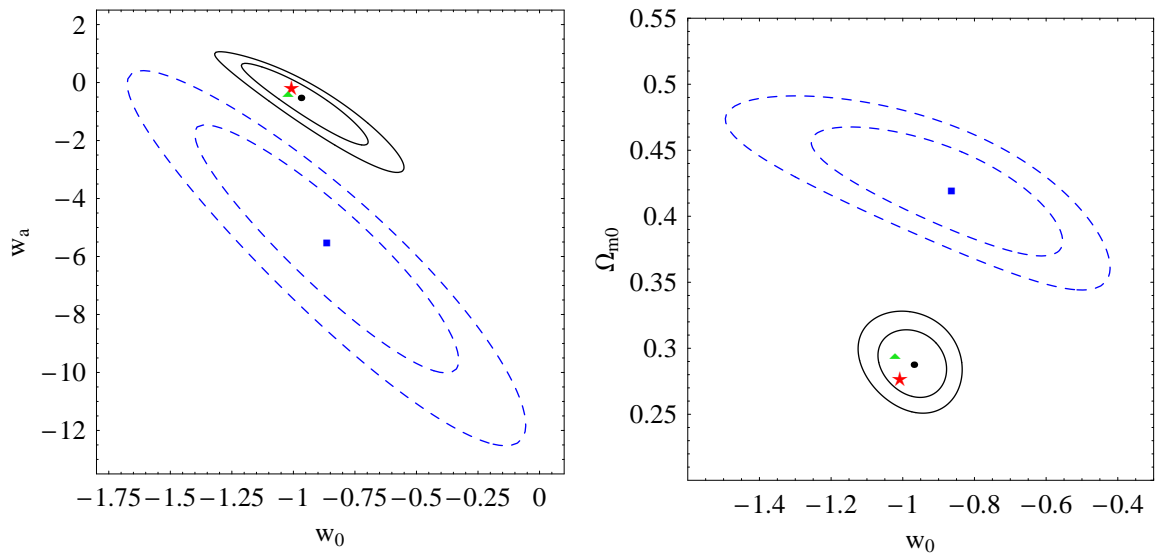}
 \caption{\label{fig1}
 The $68.3\%$ and $95.4\%$ C.L. contours in the $w_0-w_a$
 plane and the $w_0-\Omega_{m0}$ plane for the observations of
 SNIa alone (blue dashed lines) and SNIa+$A$+$R$ (black solid
 lines). We also show the best-fit values for the observations
 of SNIa alone (blue box), SNIa+$A$ (red star), SNIa+$R$ (green
 triangle) and SNIa+$A$+$R$ (black point).}
 \end{figure}
 \end{center}


\vspace{-12mm}  


\section{Data analysis}\label{sec3}

In this section, we perform the data analysis by using the
 cosmological observations mentioned in the previous section.
 In particular, we discuss the tension between CMB and SNIa in
 Sec.~\ref{subsec3}; we study the age problem in dark energy
 models in Sec~\ref{subsec4}; and the cosmological constraints
 on dark energy models are considered in Sec.~\ref{subsec5}.

\vspace{-3mm}  


\subsection{Tension between CMB and SNIa}\label{subsec3}

There is a long history of the tension between CMB and SNIa.
 In~\cite{r27,r28}, the Gold04 SNIa dataset~\cite{r29} has
 been shown to be in $2\sigma$ tension with the SNLS SNIa
 dataset~\cite{r30} and the WMAP observations. Although the
 Gold04 sample was updated to Gold06~\cite{r31} several years
 later, the tension still persisted. In~\cite{r32}, the
 tension and systematics in the Gold06 SNIa dataset has been
 investigated in great detail. Later, the number of SNIa
 has been significantly increased to $300\sim 400$ when the
 Union~\cite{r33} and Constitution~\cite{r34} SNIa datasets
 have been released. Subsequently, in~\cite{r35} it is found
 that there is still significant tension between the WMAP CMB
 data and the Union/Constitution SNIa datasets. The latest
 SNIa dataset is the Union2 compiled in~\cite{r20}, which is
 an updated version of Union. However, in~\cite{r36,r37} it
 is found that the tension between the WMAP CMB data and the
 Union2 SNIa dataset still persisted.

Obviously, the tension comes from both sides: CMB and SNIa.
 In~\cite{r32,r35,r36,r37}, the authors mainly concentrated
 in SNIa datasets, and tried to find the outliers responsible
 for the tension. However, from the works by
 L\&L~\cite{r8,r9,r10}, we become aware of the possible
 systematics in the CMB data. So, it is natural to see whether
 the alternative CMB data of L\&L~\cite{r8} can alleviate the
 tension.

In this subsection, we follow the discussions in~\cite{r35},
 and consider the familiar Chevallier-Polarski-Linder (CPL)
 model~\cite{r38}, in which the equation-of-state parameter
 (EoS) of dark energy is parameterized as
 \be{eq13}
 w_{de}=w_0+w_a(1-a)=w_0+w_a\frac{z}{1+z}\,,
 \ee
 where $w_0$ and $w_a$ are constants. As is well known, the
 corresponding $E(z)$ is given by~\cite{r39,r40,r41,r71,r19}
 \be{eq14}
 E(z)=\left[\Omega_{m0}(1+z)^3
 +\left(1-\Omega_{m0}\right)(1+z)^{3(1+w_0+w_a)}
 \exp\left(-\frac{3w_a z}{1+z}\right)\right]^{1/2}.
 \ee
 Here, we consider the shift parameter $R$ from
 the alternative CMB data of L\&L~\cite{r8}, the latest Union2
 SNIa dataset~\cite{r20}, and the distance parameter $A$~\cite{r24}
 with $n_s$ from L\&L~\cite{r8}. We fit the CPL model to
 the observations of SNIa alone, SNIa+$A$, SNIa+$R$, and
 SNIa+$A$+$R$, respectively. The best-fit values are presented
 in Table~\ref{tab2}. In Fig.~\ref{fig1}, we also present the
 $68.3\%$ and $95.4\%$ C.L. contours in the $w_0-w_a$ plane
 and the $w_0-\Omega_{m0}$ plane. From Fig.~\ref{fig1}, we find
 that there is still tension (beyond $2\sigma$) between the
 alternative CMB data of L\&L~\cite{r8} and the latest Union2 SNIa
 dataset~\cite{r20}. However, when we compare Fig.~\ref{fig1}
 and Table~\ref{tab2} of the present work with Figs.~1, 2 and
 Tables I, II of~\cite{r35}, it is easy to see that the tension
 has been alleviated to some extent in fact. The alleviation
 mainly comes from two sides: the best-fit $\Omega_{m0}$ for
 L\&L's alternative CMB data is increased, whereas the best-fit
 $\Omega_{m0}$ for the Union2 SNIa dataset is decreased. It is
 worth noting that the contribution to the alleviation from
 Union2 SNIa dataset is larger than the one from L\&L's
 alternative CMB data~\cite{r8}.


\subsection{Age problem}\label{subsec4}

In history, the age problem played an important role in the
 cosmology for many times (see e.g.~\cite{r42} for a brief
 review). The main idea is very simple: the universe cannot
 be younger than its constituents. For example, the
 matter-dominated Friedmann-Robertson-Walker (FRW) universe
 can be ruled out because its age is smaller than the ages
 inferred from old globular clusters. The age problem becomes
 even more serious when we consider the age of the universe at
 high redshift (rather than today, $z=0$). So far, there
 are some old high redshift objects (OHROs) been discovered.
 For instance, the 3.5~Gyr old galaxy LBDS~53W091 at redshift
 $z=1.55$~\cite{r43}, the 4.0~Gyr old galaxy LBDS~53W069 at
 redshift $z=1.43$~\cite{r44}, the 4.0~Gyr old radio galaxy
 3C~65 at $z=1.175$~\cite{r45}, and the high redshift quasar
 B1422+231 at $z=3.62$ whose best-fit age is 1.5~Gyr with a
 lower bound of 1.3~Gyr~\cite{r46}. In addition, the old
 quasar APM~08279+5255 at $z=3.91$ is also used extensively,
 whose age is estimated to be 2.0---3.0~Gyr~\cite{r47,r48}.
 To assure the robustness of our analysis, we use the most
 conservative lower age estimate 2.0~Gyr for the old quasar
 APM~08279+5255 at $z=3.91$~\cite{r47,r48}, and the lower age
 estimate 1.3~Gyr for the high redshift quasar B1422+231 at
 $z=3.62$~\cite{r46}. In the literature, there are many works
 on the age problem in the dark energy models, see
 e.g.~\cite{r42,r49,r50,r51,r52,r53,r54,r55} and references
 therein.

In this subsection, we would like to consider the age problems
 in the flat $\Lambda$CDM model and the holographic dark energy
 (HDE) model, which have been discussed in the literature
 by using the earlier observational data (see
 e.g.~\cite{r42,r50,r51,r54}). Of course, our main goal is to
 see whether the age problem can be alleviated with the
 alternative CMB data of L\&L~\cite{r8}.


 \begin{center}
 \begin{figure}[tbp]
 \centering
 \includegraphics[width=1.0\textwidth]{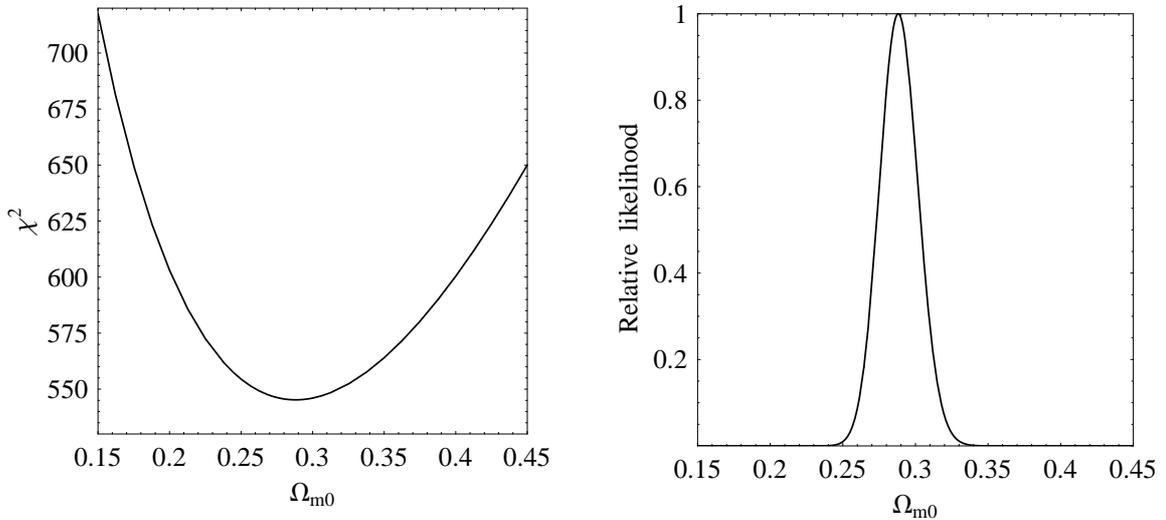}
 \caption{\label{fig2}
 The $\chi^2$ and likelihood ${\cal L}\propto e^{-\chi^2/2}$ as
 functions of $\Omega_{m0}$ for the $\Lambda$CDM model.}
 \end{figure}
 \end{center}


\vspace{-11mm}  


 \begin{table}[tbp]
 \begin{center}
 \vspace{3mm}  
 \begin{tabular}{ccccccc} \hline\hline \\[-3.5mm]
 Description & ~~~~$\Omega_{m0}$~~~~ & ~~$S(3.91)$~~
 & ~~$S(1.43)$~~ & ~~$S(1.55)$~~ & ~~$S(1.175)$~~
 & ~~$S(3.62)$~~ \\[1.2mm] \hline \\[-3.5mm]
 best fit & 0.288 & 0.799 & 1.120 & 1.195
 & 1.310 & 1.345 \\
 $1\sigma$ lower bound & 0.275 & 0.817
 & 1.144 & 1.221 & 1.337 & 1.376 \\
 $1\sigma$ upper bound & 0.302 & 0.781 & 1.097
 & 1.170 & 1.283 & 1.315 \\
 $2\sigma$ lower bound & 0.263 & 0.836 & 1.169
 & 1.248 & 1.366 & 1.408 \\
 ~~$2\sigma$ upper bound~~ & 0.316 & 0.763 & 1.074
 & 1.145 & 1.257 & 1.286
 \\[0.8mm] \hline\hline
 \end{tabular}
 \end{center}
 \caption{\label{tab3} The ratio $S(z)\equiv T_z(z)/T_{obj}$
 at $z=3.91$, $1.43$, $1.55$, $1.175$ and $3.62$, for various
 model parameters $\Omega_{m0}$ (within $2\sigma$ uncertainty)
 of the $\Lambda$CDM model (the corresponding $h=0.696$).}
 \end{table}



\subsubsection{Age problem in the $\Lambda$CDM model}\label{subsec4a}

The age of our universe at redshift $z$ is given
 by~\cite{r42,r50,r54}
 \be{eq15}
 t(z)=
 \int_z^\infty\frac{d\tilde{z}}{(1+\tilde{z})H(\tilde{z})}\,.
 \ee
 It is convenient to introduce the so-called dimensionless
 age parameter~\cite{r50,r54}
 \be{eq16}
 T_z(z)\equiv H_0 t(z)=\int_z^\infty
 \frac{d\tilde{z}}{(1+\tilde{z})E(\tilde{z})}\,.
 \ee
 At any redshift, the age of our universe should be larger
 than, at least equal to, the age of the OHRO, namely
 \be{eq17}
 T_z(z)\geq T_{obj}\equiv H_0 t_{obj}\,,
 ~~~~~~~{\rm or~equivalently,}~~~~~~~
 S(z)\equiv T_z(z)/T_{obj}\geq 1\,,
 \ee
 where $t_{obj}$ is the age of the OHRO. It is worth noting
 that from Eq.~(\ref{eq16}), $T_z(z)$ is independent of the
 Hubble constant $H_0$. On the other hand, from
 Eqs.~(\ref{eq17}), $T_{obj}$ is proportional to the Hubble
 constant $H_0$. The lower $H_0$, the smaller $T_{obj}$ is.


 \begin{center}
 \begin{figure}[tbp]
 \centering
 \includegraphics[width=0.5\textwidth]{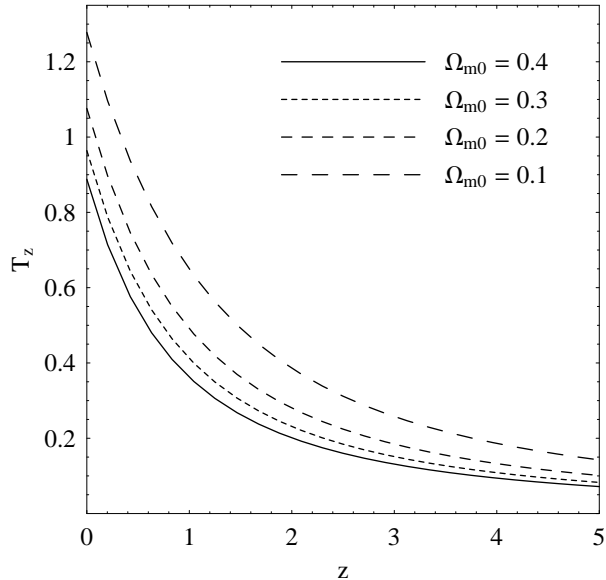}
 \caption{\label{fig3}
 The dimensionless age parameter $T_z$ as a function of redshift $z$
 for various $\Omega_{m0}$ in the $\Lambda$CDM model.}
 \end{figure}
 \end{center}


\vspace{-7mm}  

As is well known, for the flat $\Lambda$CDM model,
 \be{eq18}
 E(z)=\sqrt{\Omega_{m0}(1+z)^3+(1-\Omega_{m0})}\,.
 \ee
 It is easy to obtain the total
 $\chi^2=\tilde{\chi}^2_\mu+\chi^2_{\rm CMB}+\chi^2_{\rm BAO}$
 as a function of the single model parameter $\Omega_{m0}$ for
 the $\Lambda$CDM model, by using the combined observations of
 the shift parameter $R$ from the alternative CMB data of
 L\&L~\cite{r8}, the latest Union2 SNIa dataset~\cite{r20},
 and the distance parameter $A$~\cite{r24} with $n_s$ from
 L\&L~\cite{r8}. We present the corresponding $\chi^2$ and
 likelihood ${\cal L}\propto e^{-\chi^2/2}$ in Fig.~\ref{fig2}.
 The best fit has $\chi^2_{min}=545.239$, whereas the best-fit
 parameter is $\Omega_{m0}=0.288^{+0.013}_{-0.013}$ (with
 $1\sigma$ uncertainty) $^{+0.027}_{-0.026}$ (with $2\sigma$
 uncertainty). The corresponding $h=0.696$ for the best fit.
 It is worth noting that although the best-fit $\Omega_{m0}$
 is smaller than the one from the alternative CMB data of L\&L
 alone~\cite{r8} (see also Table~\ref{tab1}) due to the
 influence of SNIa and BAO, it is still larger than the one
 from WMAP alone~\cite{r14,r15} (see also Table~\ref{tab1})
 and the one from WMAP+SNIa+BAO~\cite{r14,r15}.

In Table~\ref{tab3}, we present the ratio
 $S(z)\equiv T_z(z)/T_{obj}$ at $z=3.91$, $1.43$, $1.55$,
 $1.175$ and $3.62$, for various model parameters $\Omega_{m0}$
 (within $2\sigma$ uncertainty) of the $\Lambda$CDM model.
 Obviously, $T_z(z)> T_{obj}$ holds at $z=1.43$, $1.55$,
 $1.175$ and $3.62$, whereas $T_z(z)< T_{obj}$ at $z=3.91$.
 So, the old quasar APM~08279+5255 at $z=3.91$ cannot be
 accommodated (beyond $2\sigma$). The age problem still
 exists in the $\Lambda$CDM model, even when the alternative
 CMB data of L\&L~\cite{r8} has been taken into account.

Let us have a closer observation. As mentioned above, the
 alternative CMB data of L\&L~\cite{r8} favor a
 larger $\Omega_{m0}$ (this point has also been mentioned by
 L\&L themselves~\cite{r8}). So, it is natural to see how the
 model parameter $\Omega_{m0}$ affects the age of our universe
 at redshift $z$. In Fig.~\ref{fig3}, we present
 the dimensionless age parameter $T_z$ as a function of
 redshift $z$ for various $\Omega_{m0}$ in the $\Lambda$CDM
 model (notice that from Eq.~(\ref{eq16}), $T_z(z)$ is
 independent of the Hubble constant $H_0$). It is easy to see
 that at any redshift $z$, the larger $\Omega_{m0}$, the
 smaller age is. Therefore, the age problem becomes even worse
 in the $\Lambda$CDM model, since the alternative CMB data of
 L\&L~\cite{r8} favor a larger $\Omega_{m0}$.


\subsubsection{Age problem in the HDE model}\label{subsec4b}

Previously, in~\cite{r54}, the age problem in the HDE model
 has been discussed in great detail. Here we would like to
 see whether the age problem can be alleviated with the
 alternative CMB data of L\&L~\cite{r8}.


 \begin{center}
 \begin{figure}[tbhp]
 \centering
 \includegraphics[width=0.5\textwidth]{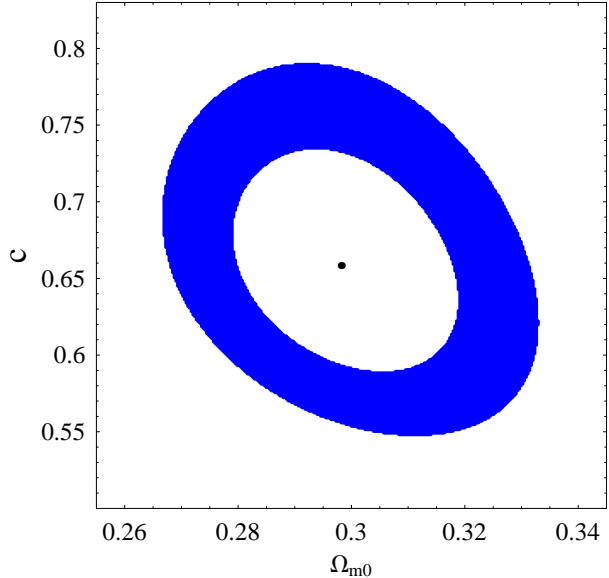}
 \caption{\label{fig4}
 The $68.3\%$ and $95.4\%$ C.L. contours in the
 $\Omega_{m0}-c$ parameter space for the HDE model. The
 best-fit parameters are also indicated by a black solid point.}
 \end{figure}
 \end{center}


\vspace{-7mm} 

The HDE model~\cite{r56,r57} was proposed from the holographic
 principle~\cite{r58} in the string theory. HDE is now an
 interesting candidate of dark energy, which has been studied
 extensively in the literature. The energy density of HDE
 reads~\cite{r57} (see also e.g.~\cite{r54,r59,r60})
 \be{eq19}
 \rho_\Lambda=3c^2 m_p^2 L^{-2}\,,
 \ee
 where the numerical constant $3c^2$ is introduced for
 convenience, and $m_p\equiv (8\pi G)^{-1/2}$ is the reduced
 Planck mass. In~\cite{r57}, $L$ has been chosen to be the
 future event horizon
 \be{eq20}
 R_h=a\int_t^\infty\frac{d\tilde{t}}{a}
 =a\int_a^\infty\frac{d\tilde{a}}{H\tilde{a}^2}\,.
 \ee
 From Eqs.~(\ref{eq19}), (\ref{eq20}), and the energy conservation
 equation $\dot{\rho}_\Lambda+3H\rho_\Lambda(1+w_\Lambda)=0$, it is
 easy to find that (see e.g.~\cite{r54,r57,r59,r60})
 \be{eq21}
 \frac{d\Omega_\Lambda}{dz}=-(1+z)^{-1}\Omega_\Lambda
 (1-\Omega_\Lambda)\left(1+\frac{2}{c}\sqrt{\Omega_\Lambda}\right),
 \ee
 where $\Omega_\Lambda$ is the fractional energy density of HDE.
 From the Friedmann equation
 $H^2=\left(\rho_m+\rho_\Lambda\right)/\left(3m_p^2\right)$, we have
 \be{eq22}
 E(z)= \left[\frac{\Omega_{m0}(1+z)^3}{1-\Omega_\Lambda(z)}
 \right]^{1/2}\,.
 \ee
 There are 2 independent model parameters, namely $\Omega_{m0}$
 and $c\,$. One can obtain $\Omega_\Lambda(z)$ by solving the
 differential equation~(\ref{eq21}) with the initial condition
 $\Omega_\Lambda(z=0)=1-\Omega_{m0}$. Substituting
 $\Omega_\Lambda(z)$ into Eq.~(\ref{eq22}), we can find
 the corresponding $E(z)$ and then the total
 $\chi^2=\tilde{\chi}^2_\mu+\chi^2_{\rm CMB}+\chi^2_{\rm BAO}$.
 By minimizing the total $\chi^2$, we find the
 best-fit parameters $\Omega_{m0}=0.298$ and $c=0.658$,
 while $\chi^2_{min}=546.628$. The corresponding $h=0.703$ for
 the best fit. In Fig.~\ref{fig4}, we present the corresponding
 $68.3\%$ and $95.4\%$ C.L. contours in the $\Omega_{m0}-c$
 parameter space for the HDE model. Comparing with the results
 of~\cite{r59,r60}, it is easy to see that $\Omega_{m0}$
 becomes larger, whereas $c$ becomes smaller. Note that $c<1$
 within $2\sigma$ region, the EoS of HDE can be less than
 $-1$~\cite{r61}.


 \begin{table}[tbhp]
 \begin{center}
 \begin{tabular}{ccccccc} \hline\hline \\[-3.5mm]
 Description & $(\Omega_{m0},\,c)$ & ~~$S(3.91)$~~
 & ~~$S(1.43)$~~ & ~~$S(1.55)$~~ & ~~$S(1.175)$~~
 & ~~$S(3.62)$~~ \\[1.2mm] \hline \\[-3.5mm]
 best fit & $(0.298,\,0.658)$ & 0.771 & 1.079
 & 1.151 & 1.263 & 1.298 \\
 $1\sigma$ left edge & $(0.279,\,0.668)$ & 0.797 & 1.112
 & 1.187 & 1.301 & 1.341 \\
 $1\sigma$ right edge & $(0.319,\,0.642)$ & 0.746 & 1.046
 & 1.116 & 1.226 & 1.257 \\
 $1\sigma$ bottom edge & $(0.303,\,0.589)$ & 0.766 & 1.076
 & 1.147 & 1.261 & 1.290 \\
 $1\sigma$ top edge & $(0.296,\,0.734)$ & 0.772 & 1.076
 & 1.148 & 1.259 & 1.300 \\
 $2\sigma$ left edge & $(0.267,\,0.678)$ & 0.814 & 1.135
 & 1.211 & 1.327 & 1.370 \\
 $2\sigma$ right edge & $(0.333,0.622)$ & 0.731 & 1.027
 & 1.095 & 1.204 & 1.231 \\
 ~~$2\sigma$ bottom edge~~ & ~~~~$(0.308,\,0.548)$~~~~ & 0.761
 & 1.070 & 1.141 & 1.255 & 1.281 \\
 $2\sigma$ top edge & $(0.294,\,0.789)$ & 0.774 & 1.075
 & 1.148 & 1.257 & 1.302
 \\[0.8mm] \hline\hline
 \end{tabular}
 \end{center}
 \caption{\label{tab4} The ratio $S(z)\equiv T_z(z)/T_{obj}$
 at $z=3.91$, $1.43$, $1.55$, $1.175$ and $3.62$, for various
 model parameters $(\Omega_{m0},\,c)$ (within $2\sigma$
 uncertainty) of the HDE model (the corresponding $h=0.703$).}
 \end{table}



 \begin{center}
 \begin{figure}[tbhp]
 \centering
 \vspace{4mm} 
 \includegraphics[width=0.49\textwidth]{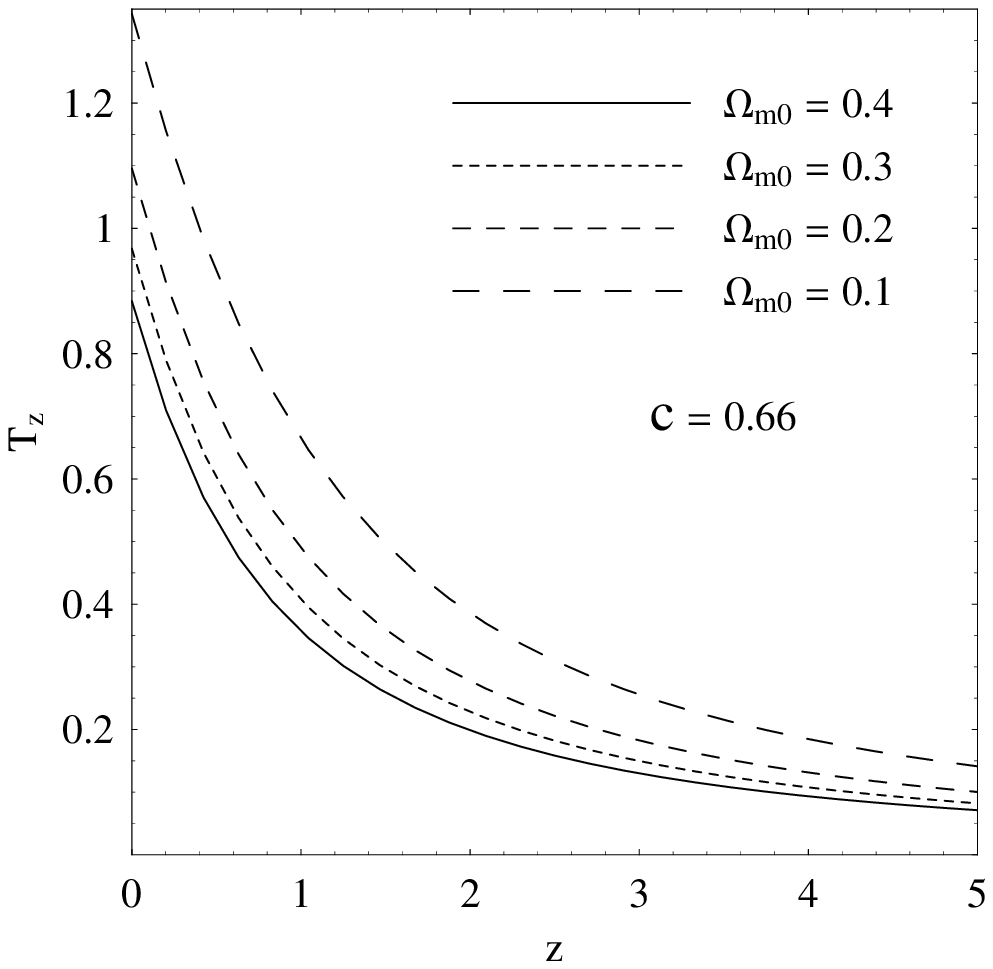}\hfill
 \includegraphics[width=0.49\textwidth]{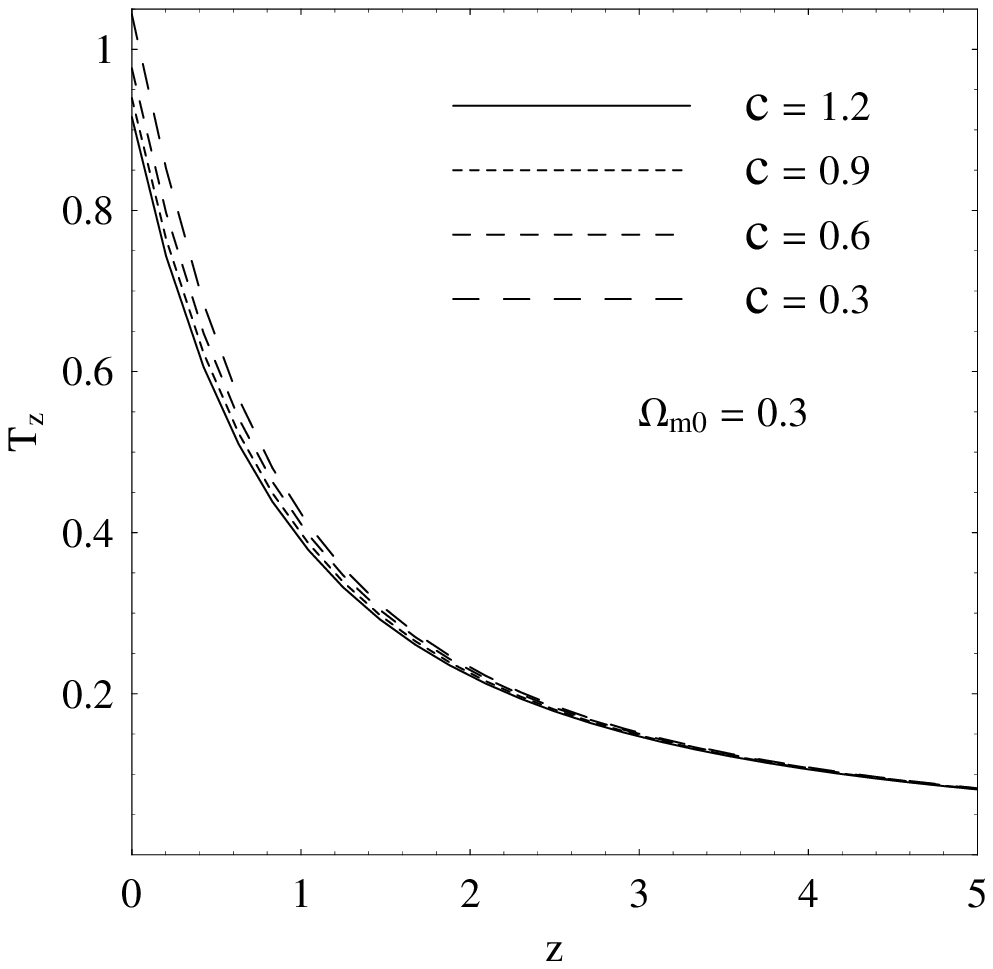}
 \caption{\label{fig5}
 The dimensionless age parameter $T_z$ as a function of redshift $z$
 for various $\Omega_{m0}$ and $c$ in the HDE model.}
 \end{figure}
 \end{center}


\vspace{-7mm} 

In Table~\ref{tab4}, we present the ratio
 $S(z)\equiv T_z(z)/T_{obj}$ at $z=3.91$, $1.43$, $1.55$,
 $1.175$ and $3.62$, for various model parameters
 $(\Omega_{m0},\,c)$ (within $2\sigma$ uncertainty) of the HDE
 model. Obviously, $T_z(z)> T_{obj}$ holds at $z=1.43$, $1.55$,
 $1.175$ and $3.62$, whereas $T_z(z)< T_{obj}$ at $z=3.91$.
 So, the old quasar APM~08279+5255 at $z=3.91$ cannot be
 accommodated (beyond $2\sigma$). The age problem still exists
 in the HDE model, even when the alternative CMB data of
 L\&L~\cite{r8} has been taken into account. Further, comparing
 with the results in~\cite{r54}, it is easy to find that
 actually the age problem becomes even worse in the HDE model.

Again, we would like to see how the model parameters $\Omega_{m0}$
 and $c$ affect the age of our universe at redshift $z$.
 In Fig.~\ref{fig5}, we present the dimensionless age parameter
 $T_z$ as a function of redshift $z$ for various $\Omega_{m0}$
 and $c$ in the HDE model (notice that from Eq.~(\ref{eq16}),
 $T_z(z)$ is independent of the Hubble constant~$H_0$). It is
 easy to see that at any redshift $z$, for a fixed $c\,$, the
 larger $\Omega_{m0}$, the smaller age is; on the other hand,
 for a fixed $\Omega_{m0}$, the larger $c\,$, the smaller
 age is. However, from Fig.~\ref{fig5}, we also find that
 the influence to the age from $\Omega_{m0}$ is significantly
 stronger than the one from $c\,$. As mentioned above,
 the alternative CMB data of L\&L~\cite{r8} favor a larger
 $\Omega_{m0}$ and a smaller $c\,$. Therefore, it is not
 surprising to find that the age problem becomes even worse
 in the HDE model.


 \begin{center}
 \begin{figure}[tbhp]
 \centering
 \includegraphics[width=0.5\textwidth]{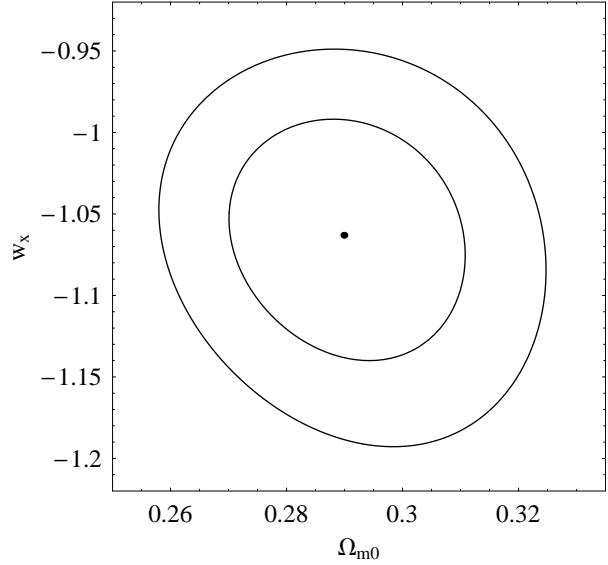}
 \caption{\label{fig6}
 The $68.3\%$ and $95.4\%$ C.L. contours in the
 $\Omega_{m0}-w_{_X}$ parameter space for the XCDM model.
 The best-fit parameters are also indicated by a black
 solid point.}
 \end{figure}
 \end{center}


\vspace{-12mm} 


 \begin{center}
 \begin{figure}[tbhp]
 \centering
 \includegraphics[width=1.0\textwidth]{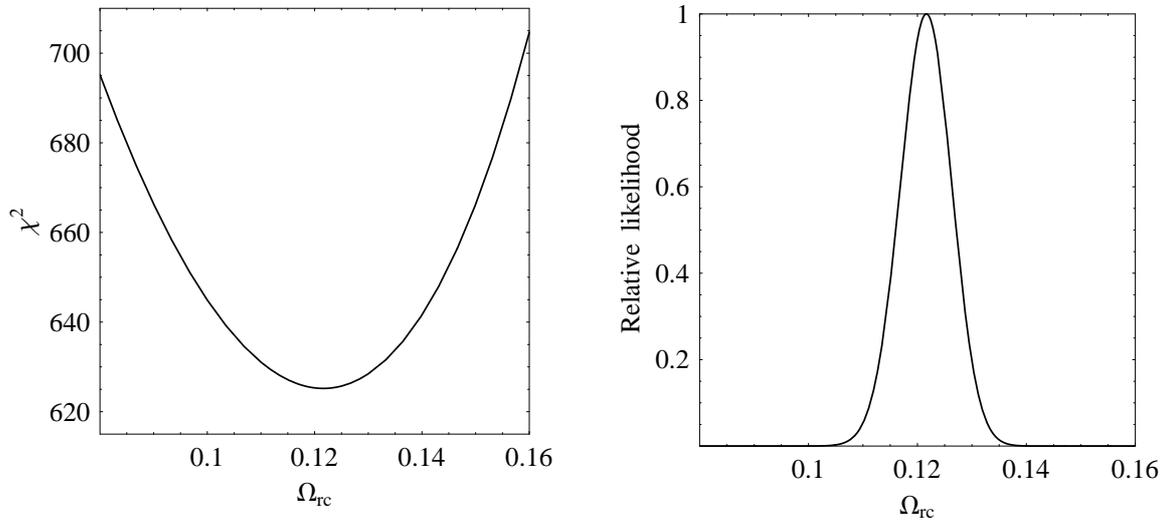}
 \caption{\label{fig7}
 The $\chi^2$ and likelihood ${\cal L}\propto e^{-\chi^2/2}$ as
 functions of $\Omega_{rc}$ for the DGP model.}
 \end{figure}
 \end{center}


\vspace{-12mm} 


\subsection{Cosmological constraints on dark energy models}\label{subsec5}

In this subsection, we consider the cosmological constraints on
 various dark energy models, by using the combined observations
 of the shift parameter $R$ from the alternative CMB data of
 L\&L~\cite{r8}, the latest Union2 SNIa dataset~\cite{r20},
 and the distance parameter $A$~\cite{r24} with $n_s$ from
 L\&L~\cite{r8}. It is worth noting that the cosmological
 constraints on $\Lambda$CDM model, CPL model and HDE model
 have already been obtained in Secs.~\ref{subsec3}
 and~\ref{subsec4}. In the followings, we further consider the
 cosmological constraints on the other four dark energy models.


\subsubsection{XCDM model}\label{subsec5a}

It is well known that in the spatially flat universe which
 contains pressureless matter and dark energy whose EoS
 is a constant $w_{_X}$, the corresponding $E(z)$ is given
 by (see e.g.~\cite{r59,r60})
 \be{eq23}
 E(z)=\sqrt{\Omega_{m0}(1+z)^3+
 (1-\Omega_{m0})(1+z)^{3(1+w_{_X})}}\,.
 \ee
 By minimizing the corresponding total
 $\chi^2=\tilde{\chi}^2_\mu+\chi^2_{\rm CMB}+\chi^2_{\rm BAO}$,
 we find the best-fit parameters $\Omega_{m0}=0.290$ and
 $w_{_X}=-1.063$, while $\chi^2_{min}=543.45$. The
 corresponding $h=0.700$ for the best fit. In Fig.~\ref{fig6},
 we present the corresponding $68.3\%$ and $95.4\%$ C.L.
 contours in the $\Omega_{m0}-w_{_X}$ parameter space for
 the XCDM model. Comparing with the corresponding results
 obtained in~\cite{r59,r60}, it is easy to find
 that $\Omega_{m0}$ becomes fairly larger, whereas $w_{_X}$
 becomes slightly smaller.


\subsubsection{DGP model}\label{subsec5b}

One of the simplest modified gravity models is the well-known
 Dvali-Gabadadze-Porrati~(DGP) braneworld model~\cite{r62,r63},
 which entails altering the Einstein-Hilbert action by a term
 arising from large extra dimensions. For a list of references
 on the DGP model, see e.g.~\cite{r64,r65} and references
 therein. As is well known, for the spatially flat DGP model
 (here we only consider the self-accelerating branch), $E(z)$
 is given by~\cite{r63,r64,r65}
 \be{eq24}
 E(z)=\sqrt{\Omega_{m0}(1+z)^3+\Omega_{rc}}+
 \sqrt{\Omega_{rc}}\,,
 \ee
 where $\Omega_{rc}$ is a constant. It is easy to see that
 $E(z=0)=1$ requires
 \be{eq25}
 \Omega_{m0}=1-2\sqrt{\Omega_{rc}}\,.
 \ee
 Therefore, the DGP model has only one independent
 model parameter $\Omega_{rc}$. Notice that
 $0\leq\Omega_{rc}\leq 1/4$ is required by
 $0\leq\Omega_{m0}\leq 1$. It is easy to obtain the total
 $\chi^2=\tilde{\chi}^2_\mu+\chi^2_{\rm CMB}+\chi^2_{\rm BAO}$
 as a function of the single model parameter $\Omega_{rc}$.
 In Fig.~\ref{fig7}, we plot the corresponding $\chi^2$ and
 likelihood ${\cal L}\propto e^{-\chi^2/2}$. The
 best fit has $\chi^2_{min}=625.2$, whereas the best-fit
 parameter is $\Omega_{rc}=0.122^{+0.005}_{-0.005}$ (with
 $1\sigma$ uncertainty) $^{+0.009}_{-0.010}$ (with $2\sigma$
 uncertainty). The corresponding $h=0.678$ for the best fit.
 From Eq.~(\ref{eq25}), $\Omega_{m0}=0.302$ corresponds to
 the best-fit $\Omega_{rc}$. Comparing with the corresponding
 results obtained in~\cite{r59,r60}, it is easy to find that
 $\Omega_{rc}$ becomes fairly smaller, and hence $\Omega_{m0}$
 becomes fairly larger.


\subsubsection{New agegraphic dark energy model}\label{subsec5c}

\begin{center}
 \begin{figure}[tbp]
 \centering
 \includegraphics[width=1.0\textwidth]{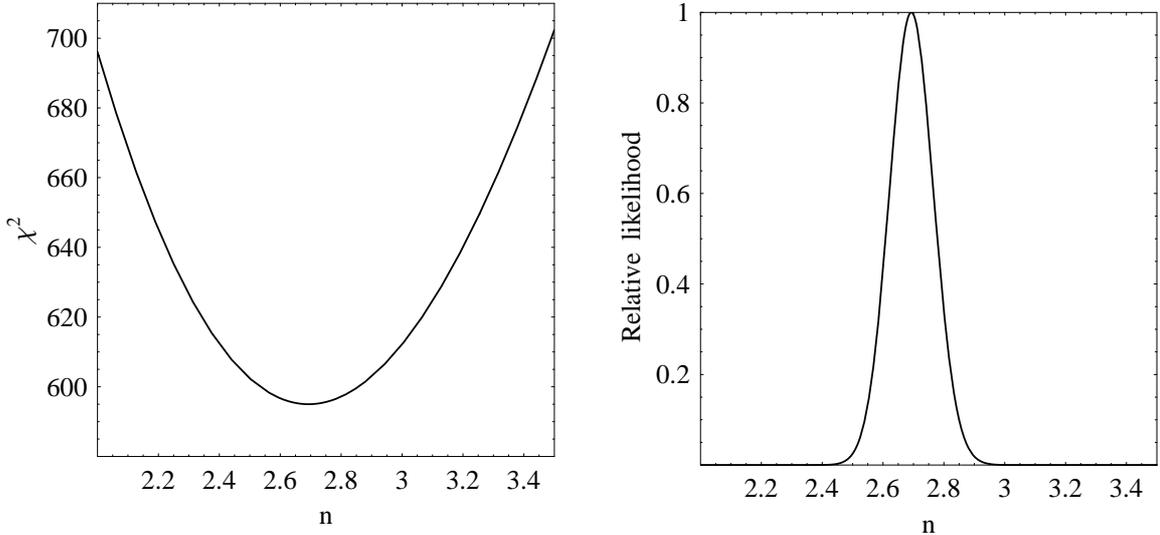}
 \caption{\label{fig8}
 The $\chi^2$ and likelihood ${\cal L}\propto e^{-\chi^2/2}$ as
 functions of $n$ for the NADE model.}
 \end{figure}
 \end{center}

The so-called new agegraphic dark energy (NADE) model was
 proposed in~\cite{r66,r67}, based on the K\'{a}rolyh\'{a}zy
 uncertainty relation which arises from quantum mechanics
 together with general relativity. In fact, the NADE model
 is an upgraded version of the agegraphic dark energy (ADE)
 model~\cite{r68,r69,r70}. In the NADE model, the dark energy
 density is given by~\cite{r66,r67}
 \be{eq26}
 \rho_q=\frac{3n^2m_p^2}{\eta^2}\,,
 \ee
 where $n$ is a constant of order unity; $\eta$ is
 the conformal time which is defined by
 \be{eq27}
 \eta\equiv\int\frac{dt}{a}=\int\frac{da}{a^2H}\,.
 \ee
 From the Friedmann equation
 $H^2=\left(\rho_m+\rho_q\right)/\left(3m_p^2\right)$,
 the energy conservation equation $\dot{\rho}_m+3H\rho_m=0$,
 Eqs.~(\ref{eq26}) and (\ref{eq27}), it is easy to find
 that the equation of motion for $\Omega_q$, the fractional
 energy density of NADE, is given by~\cite{r66,r67}
 \be{eq28}
 \frac{d\Omega_q}{dz}=-\Omega_q\left(1-\Omega_q\right)
 \left[3(1+z)^{-1}-\frac{2}{n}\sqrt{\Omega_q}\right].
 \ee
 On the other hand, from the energy conservation
 equation $\dot{\rho}_q+3H(\rho_q+p_q)=0$,
 Eqs.~(\ref{eq26}) and (\ref{eq27}), one can find that
 the EoS of NADE is given by~\cite{r66,r67}
 \be{eq29}
 w_q\equiv\frac{p_q}{\rho_q}=
 -1+\frac{2}{3n}\frac{\sqrt{\Omega_q}}{a}\,.
 \ee
 From Eqs.~(\ref{eq26}), (\ref{eq27}) and (\ref{eq29}), we
 have a very important result, namely, $\Omega_q=n^2a^2/4$
 in the matter-dominated epoch (we strongly refer
 to~\cite{r66} for detailed arguments). Thanks to this special
 analytic feature $\Omega_q=n^2a^2/4=n^2(1+z)^{-2}/4$ in the
 matter-dominated epoch, NADE is a single-parameter model in
 practice. If $n$ is given, we can obtain $\Omega_q(z)$ from
 Eq.~(\ref{eq28}) with the initial condition
 $\Omega_q(z_{ini})=n^2(1+z_{ini})^{-2}/4$ at any $z_{ini}$
 which is deep enough into the matter-dominated epoch (we
 choose $z_{ini}=2000$ as in~\cite{r66}), instead of
 $\Omega_q(z=0)=1-\Omega_{m0}$ at $z=0$. Then, all other
 physical quantities, such as $\Omega_m(z)=1-\Omega_q(z)$
 and $w_q(z)$ in Eq.~(\ref{eq29}), can be obtained correspondingly.
 So, $\Omega_{m0}=\Omega_m(z=0)$, $\Omega_{q0}=\Omega_q(z=0)$ and
 $w_{q0}=w_q(z=0)$ are {\em not} independent model parameters.
 The only free model parameter is $n$ in the NADE model.
 From the Friedmann equation
 $H^2=\left(\rho_m+\rho_q\right)/\left(3m_p^2\right)$, we have
 \be{eq30}
 E(z)=\left[\frac{\Omega_{m0}(1+z)^3}{1-\Omega_q(z)}\right]^{1/2}.
 \ee
 If the single model parameter $n$ is given, we can obtain
 $\Omega_q(z)$ from Eq.~(\ref{eq28}) with the initial condition
 $\Omega_q(z_{ini})=n^2(1+z_{ini})^{-2}/4$ at $z_{ini}$. Thus,
 we get $\Omega_{m0}=1-\Omega_q(z=0)$. Then, $E(z)$ is on hand.
 Therefore, we can find the corresponding total
 $\chi^2=\tilde{\chi}^2_\mu+\chi^2_{\rm CMB}+\chi^2_{\rm BAO}$.
 In Fig.~\ref{fig8}, we plot the total
 $\chi^2$ and likelihood ${\cal L}\propto e^{-\chi^2/2}$ as
 functions of $n$. The best fit has $\chi^2_{min}=595.001$,
 whereas the best-fit parameter is $n=2.693^{+0.073}_{-0.072}$
 (with $1\sigma$ uncertainty) $^{+0.146}_{-0.143}$ (with
 $2\sigma$ uncertainty). The corresponding $h=0.681$ for the
 best fit. On the other hand, we find that $\Omega_{m0}=0.299$,
 $\Omega_{q0}=0.701$ and $w_{q0}=-0.793$ correspond to the
 best-fit $n$. Comparing with the corresponding results
 obtained in~\cite{r59,r60,r71}, we find that $n$ becomes
 fairly smaller, and correspondingly $\Omega_{m0}$ becomes
 fairly larger.


\vspace{-11mm} 


\subsubsection{Ricci dark energy model}\label{subsec5d}

\begin{center}
 \begin{figure}[tbp]
 \centering
 \includegraphics[width=0.5\textwidth]{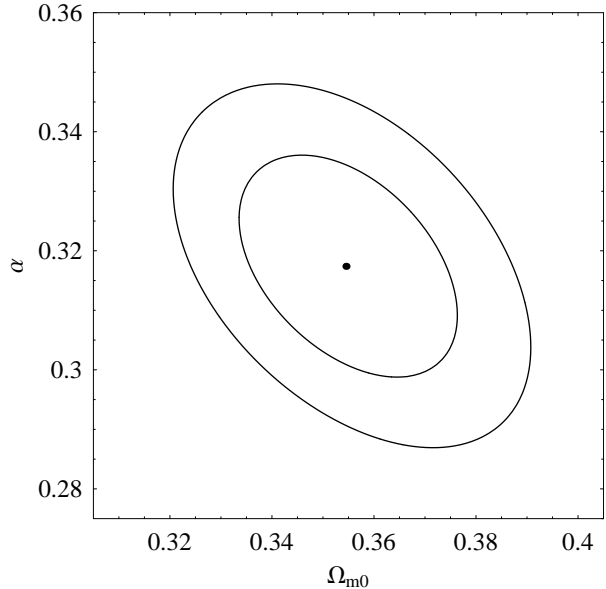}
 \caption{\label{fig9}
 The $68.3\%$ and $95.4\%$ C.L. contours in the
 $\Omega_{m0}-\alpha$ parameter space for the RDE model.
 The best-fit parameters are also indicated by a black
 solid point.}
 \end{figure}
 \end{center}

The so-called Ricci dark energy (RDE) model was proposed
 in~\cite{r72}, which can be regarded as a variant of the
 HDE model mentioned above, while its corresponding cut-off
 $L$ in Eq.~(\ref{eq19}) is chosen to be proportional to the
 Ricci scalar curvature radius. In~\cite{r72}, there is no
 physical motivation to this proposal for $L$ in fact. Later,
 in~\cite{r73} it is found that the Jeans length $R_{\rm CC}$
 which is determined by $R_{\rm CC}^{-2}=\dot{H}+2H^2$ gives
 the causal connection scale of perturbations in the flat
 universe. Since the Ricci scalar is also proportional to
 $\dot{H}+2H^2$ in the flat universe, the physical motivation
 for RDE has been found in~\cite{r73} actually. In the RDE
 model, the corresponding $\rho_\Lambda$ is given
 by~\cite{r72}
 \be{eq31}
 \rho_\Lambda=3\alpha m_p^2 \left(\dot{H}+2H^2\right),
 \ee
 where $\alpha$ is a positive constant (when $L$ is chosen
 to be $R_{\rm CC}$ in Eq.~(\ref{eq19}), one can see that
 $\alpha=c^2$ in fact). Substituting Eq.~(\ref{eq31}) into
 Friedmann equation, it is easy to find that~\cite{r72,r59,r60}
 \be{eq32}
 E(z)=\left[\,\frac{2\Omega_{m0}}{\,2-\alpha}\,(1+z)^3+
 \left(1-\frac{2\Omega_{m0}}{\,2-\alpha}\right)
 (1+z)^{4-2/\alpha}\right]^{1/2}.
 \ee
 There are two independent model parameters,
 namely $\Omega_{m0}$ and $\alpha\,$. By minimizing the
 corresponding total
 $\chi^2=\tilde{\chi}^2_\mu+\chi^2_{\rm CMB}+\chi^2_{\rm BAO}$,
 we find the best-fit parameters
 $\Omega_{m0}=0.355$ and $\alpha=0.317$, while
 $\chi^2_{min}=588.303$. In Fig.~\ref{fig9}, we present the
 corresponding $68.3\%$ and $95.4\%$ C.L. contours in the
 $\Omega_{m0}-\alpha$ parameter space for the RDE model.
 Comparing with the corresponding results obtained
 in~\cite{r59,r60}, it is easy to find that $\Omega_{m0}$
 becomes fairly larger, whereas $\alpha$ becomes significantly
 smaller.


\vspace{-10mm} 


\subsubsection{Comparison of dark energy models}\label{subsec5e}

In Table~\ref{tab5}, we summarize all the 7 dark energy
 models considered in this work. As mentioned above, it is
 common that L\&L's alternative CMB data~\cite{r8} favor
 a larger $\Omega_{m0}$ in all these dark energy models.
 Here we briefly consider the comparison of these models.
 Following~\cite{r59,r60}, we adopt three criterions used
 extensively in the literature, i.e., $\chi^2_{min}/dof$,
 Bayesian Information Criterion (BIC) and Akaike Information
 Criterion (AIC). Note that the degree of freedom $dof=N-k$,
 whereas $N$ and $k$ are the number of data points and the
 number of free model parameters, respectively. The BIC is
 defined by~\cite{r74}
 \be{eq33}
 {\rm BIC}=-2\ln{\cal L}_{max}+k\ln N\,,
 \ee
 where ${\cal L}_{max}$ is the maximum likelihood. In the
 Gaussian cases, $\chi^2_{min}=-2\ln{\cal L}_{max}$. So, the
 difference in BIC between two models is given by
 $\Delta{\rm BIC}=\Delta\chi^2_{min}+\Delta k \ln N$. The AIC
 is defined by~\cite{r75}
 \be{eq34}
 {\rm AIC}=-2\ln{\cal L}_{max}+2k\,.
 \ee
 The difference in AIC between two models is given by
 $\Delta{\rm AIC}=\Delta\chi^2_{min}+2\Delta k$.
 In Table~\ref{tab5}, we present
 $\chi^2_{min}/dof$, $\Delta$BIC and $\Delta$AIC for all the
 7 models considered in this work. Notice that $\Lambda$CDM
 has been chosen to be the fiducial model when we calculate
 $\Delta$BIC and $\Delta$AIC. From Table~\ref{tab5}, it is
 easy to see that the rank of models is coincident in all
 the 3 criterions ($\chi^2_{min}/dof$, BIC and AIC). The
 $\Lambda$CDM model is still the best, whereas the DGP model
 is still the worst. Comparing with the corresponding results
 obtained in~\cite{r59,r60}, we find that the rank of these
 models has only slight change. In other words, the influence
 to the rank from L\&L's alternative CMB data~\cite{r8} is
 fairly minor.


 \begin{table}[tbp]
 \begin{center}
 \begin{tabular}{llllllll} \hline \hline \\[-3.5mm]
 Model & $\Lambda$CDM & XCDM & CPL & DGP & NADE & HDE & RDE
 \\[1.2mm] \hline \\[-3.5mm]
 Best fit & $\Omega_{m0}=0.288$~~ & $\Omega_{m0}=0.290$~~~
 & $\Omega_{m0}=0.288$~~ & $\Omega_{rc}=0.122$~~ & $n=2.693$~~
 & $\Omega_{m0}=0.298$~~ & $\Omega_{m0}=0.355$ \\
 & & $w_{_X}=-1.063$ & $w_0=-0.969$
 & & & $c=0.658$ & $\alpha=0.317$ \\
 & & & $w_a=-0.529$ & & & &
 \\[1.2mm] \hline \\[-3.5mm]
 $\chi^2_{min}$ & 545.239 & 543.45 & 542.936 & 625.2 & 595.001
 & 546.628 & 588.303 \\
 $k$ & 1 & 2 & 3 & 1 & 1 & 2 & 2\\
 $\chi^2_{min}/dof$~~ & 0.977 & 0.976 & 0.977
 & 1.120 & 1.066 & 0.981 & 1.056 \\
 $\Delta$BIC & 0 & 4.537 & 10.349 & 79.961
 & 49.762 & 7.715 & 49.390 \\
 $\Delta$AIC & 0 & 0.211 & 1.697 & 79.961
 & 49.762 & 3.389 & 45.064 \\
 Rank & 1 & 2 & $3\sim 4$ & 7
 & $5\sim 6$ & $3\sim 4$ & $5\sim 6$
 \\[1.2mm] \hline\hline
 \end{tabular}
 \end{center}
 \caption{\label{tab5} Summarizing all the 7 dark energy
 models considered in this work.}
 \end{table}



\section{Conclusion and discussions}\label{sec6}

Recently, in a series of works by L\&L~\cite{r8,r9,r10},
 they claimed that there exists a timing asynchrony of
 $-25.6\,$ms between the spacecraft attitude and radiometer
 output timestamps in the original raw WMAP time-ordered data
 (TOD). In~\cite{r8}, L\&L reprocessed the WMAP data while the
 aforementioned timing asynchrony has been corrected, and they
 obtained an alternative CMB map in which the quadrupole dropped
 to nearly zero. In the present work, we try to see the
 implications to dark energy cosmology if L\&L are right. While
 L\&L claimed that there is a bug in the WMAP pipeline which
 leads to significantly different cosmological parameters,
 an interesting question naturally arises, namely, how robust
 is the current dark energy cosmology with respect
 to systematic errors and bugs? So, in this work, we adopt the
 alternative CMB data of L\&L as a strawman to study the
 robustness of dark energy predictions.

In this work, we found that L\&L's alternative CMB data~\cite{r8}
 favor a larger $\Omega_{m0}$ in all the dark energy models.
 As a result, we found that the tension between CMB and SNIa
 can be alleviated to some extent, since SNIa dataset usually
 favors a large $\Omega_{m0}$. However, the age problem becomes
 even worse in the dark energy models, since a larger $\Omega_{m0}$
 usually leads to a smaller age of our universe at any redshift $z$.
 On the other hand, we found that L\&L's alternative CMB
 data~\cite{r8} do not significantly change the rank of dark
 energy models from the perspective of model comparison. Of
 course, it is a big advantage that the quadrupole dropped to
 nearly zero in the alternative CMB map of L\&L
 (see~\cite{r8,r9,r10}). Altogether, we consider that it is
 better to keep neutral to L\&L's findings so far. While WMAP
 has been critical in establishing a highly successful
 cosmology, it is most important that these results are
 verified by other independent experiments, such as Planck.

Finally, we stress that this work is a hypothetical study in
 fact, since the works of L\&L~\cite{r8,r9,r10} are highly
 controversial in the community. One should be aware of the
 risk that the present work is likely to be completely
 irrelevant if it turns out that the works of L\&L are indeed
 flawed. Here are some remarks to be taken serious (we thank
 the anonymous referee for pointing out these issues). Firstly,
 in fact there are many discussions on L\&L's claims in the
 community (see e.g.~\cite{r77,r78,r79}), and many people
 (including both WMAP and Planck experts) believe that L\&L are
 mistaken in their claims. Secondly, if the high-$\ell$
 spectrum really is as discrepant as claimed by L\&L, one could
 expect that many ground-based experiments would also have
 seen this effect. Thirdly, it is worth asking whether the
 timing offset was applied correctly in the actual map making
 but not in the calibration step. Up to now, to our knowledge,
 the discussions on L\&L's claims in the community are still
 unsettled, whereas L\&L are still defending their claims
 persistently in more papers (see~\cite{r80} for instance) and
 in many conferential talks around the world. We hope that this
 ongoing controversy could be finished with a firm and reliable
 conclusion in the near future.


\section*{ACKNOWLEDGEMENTS}
We thank the anonymous referee for quite expert comments and
 useful suggestions, which help us to improve this work. We
 would like to express our gratitude to Prof.~Ti-Pei~Li
 for his talk given in the period of the Annual Conference
 of Chinese Astronomical Society, Nanning, China, November
 2010. We are grateful to Prof.~Rong-Gen~Cai and
 Prof.~Shuang~Nan~Zhang for helpful discussions. We also
 thank Minzi~Feng, as well as Xiao-Peng~Ma and Hao-Yu~Qi, for
 kind help and discussions. This work was supported in part by
 NSFC under Grant No.~10905005, the Excellent Young Scholars
 Research Fund of Beijing Institute of Technology, and the
 Fundamental Research Fund of Beijing Institute of Technology.

\renewcommand{\baselinestretch}{1.1}


\end{document}